\documentclass[journal=jpcafh,layout=twocolumn,manuscript=article]{achemso}

\usepackage[version=3]{mhchem} 
\usepackage{fontenc}
\usepackage{chemformula}
\usepackage{amsmath}
\usepackage{amsfonts}
\usepackage{amssymb}
\usepackage{lineno}
\usepackage{mathtools}
\usepackage{xcolor}
\usepackage{setspace}
\usepackage{pdfpages}
\usepackage{physics}

\author{H\r{a}kon I. R\o{st}} \email{hakon.rost@uib.no}
\affiliation{Department of Physics and Technology, University of Bergen, All\'egaten 55, 5007 Bergen, Norway.}
\alsoaffiliation{Department of Physics, Norwegian University of Science and Technology (NTNU), NO-7491 Trondheim, Norway.}

\author{Simon P. Cooil}
\affiliation{Department of Physics and Centre for Materials Science and Nanotechnology, University of Oslo (UiO), Oslo 0318, Norway.}

\author{Anna Cecilie Åsland}
\affiliation{Department of Physics, Norwegian University of Science and Technology (NTNU), NO-7491 Trondheim, Norway.}

\author{Jinbang Hu}
\affiliation{Department of Physics, Norwegian University of Science and Technology (NTNU), NO-7491 Trondheim, Norway.}

\author{Ayaz Ali}
\affiliation{Department of Smart Sensor Systems, SINTEF DIGITAL, Oslo, 0373, Norway.}
\alsoaffiliation{Department of Electronic Engineering, Faculty of Engineering \& Technology, University of Sindh, Jamshoro, 76080, Pakistan.}

\author{Takashi Taniguchi}
\affiliation{International Center for Materials Nanoarchitectonics, National Institute for Materials Science, 1-1 Namiki, Tsukuba 305-0044, Japan.}

\author{Kenji Watanabe}
\affiliation{Research Center for Functional Materials, National Institute for Materials Science, 1-1 Namiki, Tsukuba 305-0044, Japan.}

\author{Branson D. Belle}
\affiliation{Department of Smart Sensor Systems, SINTEF DIGITAL, Oslo, 0373, Norway.}

\author{Bodil Holst}
\affiliation{Department of Physics and Technology, University of Bergen, All\'egaten 55, 5007 Bergen, Norway.}

\author{Jerzy T. Sadowski}
\affiliation{Center for Functional Nanomaterials, Brookhaven National Laboratory, Upton, New York, 11973, USA.}

\author{Federico Mazzola}
\affiliation{Department of Molecular Sciences and Nanosystems, Ca’ Foscari University of Venice, 30172 Venice, Italy.}
\alsoaffiliation{Istituto Officina dei Materiali, Consiglio Nazionale delle Ricerche, Trieste I-34149, Italy.}

\author{Justin W. Wells} \email{j.w.wells@fys.uio.no}
\affiliation{Department of Physics, Norwegian University of Science and Technology (NTNU), NO-7491 Trondheim, Norway.}
\alsoaffiliation{Department of Physics and Centre for Materials Science and Nanotechnology, University of Oslo (UiO), Oslo 0318, Norway.}

\title{Phonon-Mediated Quasiparticle Lifetime Renormalizations in Few-Layer Hexagonal Boron Nitride}

\abbreviations{ARPES, Many-Body effects, electron-phonon coupling}

\begin{document}



\newpage

\twocolumn[
\begin{@twocolumnfalse}
\begin{abstract}
Understanding the collective behavior of the quasiparticles in solid-state systems underpins the field of non-volatile electronics, including the opportunity to control many-body effects for well-desired physical phenomena and their applications. Hexagonal boron nitride (hBN) is a wide energy bandgap semiconductor, showing immense potential as a platform for low-dimensional device heterostructures. It is an inert dielectric used for gated devices, having a negligible orbital hybridization when placed in contact with other systems. Despite its inertness, we discover a large electron mass enhancement in few-layer hBN affecting the lifetime of the $\pi$-band states. We show that the renormalization is phonon-mediated and consistent with both single- and multiple-phonon scattering events. Our findings thus unveil a so-far unknown many-body state in a wide-bandgap insulator, having important implications for devices using hBN as one of their building blocks.

\vspace{0.15in}\noindent\textbf{Keywords}: Hexagonal Boron Nitride, Graphene Heterostructures, Many-Body Interactions, Electron-Phonon Coupling, ARPES.
\end{abstract}
\end{@twocolumnfalse}
]


\newpage

Hexagonal Boron Nitride (hBN) is an inert layered compound that has gained significant attention for its compatibility with the vast majority of low-dimensional van der Waals (vdW) materials \cite{dean2010boron,geim2013van,wang2015electronic,withers2015light,kumar2016tunability,liu2016van,vsivskins2019high,Ahmed2020interplay}. It is strikingly similar to graphene both in lateral size, crystalline structure, and Debye frequency, but due to its dissimilar sub-lattices, it hosts a wide energy band gap separating the valence and conduction bands \cite{robertson1984electronic,wirtz2003ab}. For the engineering of vdW heterostructures embedded in the form of devices, hBN has proven to be a key building block due to its large capacitive coupling and current tunneling barrier \cite{ponomarenko2011tunable,lee2011electron,britnel2012electron,vsivskins2019high,im2022capacitance}. Furthermore, its chemical inertness, large energy bandgap, and high phonon energies have made it one of the most common dielectrics for use in state-of-the-art, low-dimensional devices that require atomic-scale flatness, and negligible interface doping and scattering \cite{dean2010boron,wang2015electronic,usachov2010quasi,tonkikh2016structural,ma2022epitaxial}.

Recently, hBN has been predicted to host a strong electron-phonon coupling which can compromise the performance of hBN-derived electronic devices \cite{slotman2014phonons,thingstad2020phonon}. However, experimental proof of such couplings has so far been lacking. Herein, we investigate the many-body effects of few-layer hBN supported on graphene. We discover and quantify the predicted electron-phonon coupling, and also an additional scattering effect that, together, significantly renormalizes the hBN $\pi$-band states. The latter effect is found to be consistent with a scattering process involving multiple phonons. Thus, our findings are crucial for understanding the many-body states in hBN-based devices, and for achieving increased control over their performances.

\begin{figure}[t!]
    \centering
    \includegraphics[]{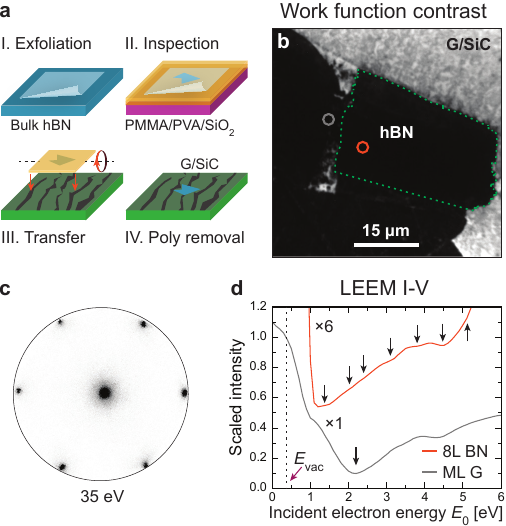}
    \caption{Exfoliated hBN on graphene/SiC. \textbf{a}: The preparation of few-layer hBN by exfoliating from a bulk crystal and transferring onto graphene/SiC using a polymer (PMMA) film. \textbf{b}: PEEM micrograph of hBN on graphene/SiC. The region encompassed by the dashed line (green) is the relevant hBN region used for further analysis. \textbf{c}: $\mu$-LEED from a $1.5~\mu$m diameter area on the hBN region marked in \textbf{b}, collected with electron energy 35~eV. \textbf{d}: LEEM I-V spectra collected from the circular areas marked in \textbf{b}. The dips in the I-V curves reveal that the hBN (orange) consists of 8 layers, while the substrate (grey) is mainly monolayer graphene. The curves have been normalized to the maximum intensity at the mirror electron microscopy energy threshold ($\approx0.4$~eV). The hBN curve has been scaled ($\times6$) and offset in intensity for improved readability.}
    \label{fig:fig1}
\end{figure}

Few (5$-$10) layer hBN flakes were exfoliated from bulk material and transferred via a polymer stack onto a substrate of epitaxial graphene on 6H-SiC(0001) (Fig.~\ref{fig:fig1}\textbf{a}). The material was then heated in ultrahigh vacuum (pressure~$\leq1\times10^{-9}$~mbar) to remove any residues of polymer from the transfer process (see the Supporting Information for details \cite{Suppl_Mat}). Using photoemission electron microscopy (PEEM), we selected a high-quality hBN flake of approximately $35\times20~\mu\text{m}^2$ lateral size (Fig.~\ref{fig:fig1}\textbf{b}). The crystalline quality across the flake was ascertained from small-area low-energy electron diffraction ($\mu$-LEED, 1.5~$\mu\text{m}$ diameter spot), showing no appreciable variation across the full flake area (Fig.~\ref{fig:fig1}\textbf{c}). The diffraction pattern taken at an electron energy of 35~eV revealed six first-order spots as expected for a stack of rotationally aligned hBN layers.

The number of vdW-bonded hBN layers in the flake was ascertained from low-energy electron reflectivity (LEER) measurements (Fig.~\ref{fig:fig1}\textbf{d}). The incident, coherent electron beam was tuned in the range 0-10~eV, and the corresponding electron reflectivity at each energy was recorded from micrographs of the surface. The averaged I--V characteristics from areas on and adjacent to the hBN flake (circles in Fig.~\ref{fig:fig1}\textbf{b}) reveal dips in the reflected intensity at energies beyond the mirror electron microscopy energy threshold at $\approx0.4$~eV. These dips represent the transmission of electrons into discrete and unoccupied states above the vacuum energy level. In the case of graphene or hBN, the minima are given by the unoccupied $\pi^{*}$-states of each atomic layer. The number of minima $n$ observed can hence be linked directly to the $n+1$ stacked layers present \cite{Ohta2008morphology,cooil2015controlling,jobst2016quantifying,raths2021growth}. From the substrate region (grey), one dominant dip ($\approx2.1$~eV) can be observed between the higher intensity regions (0.5 and 6.0~eV), suggesting the substrate is mainly monolayer graphene on top of $(6\sqrt{3}\times6\sqrt{3})\text{R}30^{\circ}$-reconstructed SiC(0001) \cite{Ohta2008morphology}. In comparison, the hBN flake (orange) shows 7 dips originating from 8 stacked layers \cite{jobst2016quantifying}. For the electronic structure measurements, we will refer to this 8-layer region from now on.

\begin{figure}[t!]
    \centering
    \includegraphics[]
    {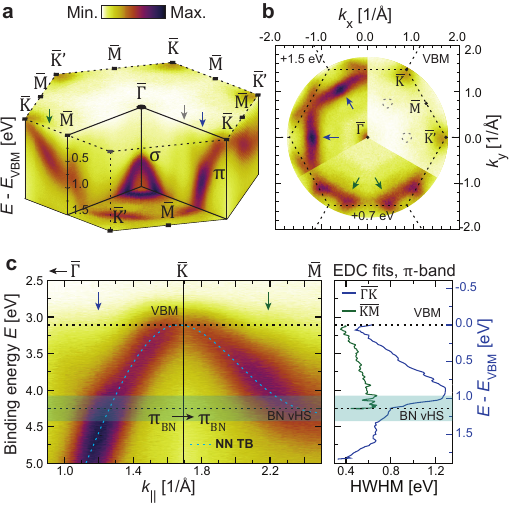}
    \caption{The electronic structure of hBN on graphene/SiC. \textbf{a}: A volumetric 3D rendering of the hBN electronic structure within the 1st BZ. A binding energy range of 0-1.5~eV is shown relative to the maximum of the valence ($\pi$) band. Prominent scattering interactions have been marked by arrows. \textbf{b}: A sequence of constant energy ($k_{\text{x}}$~\emph{vs.}~$k_{\text{y}}$) surfaces at binding energies relative to the valence band maximum (VBM). Each surface is shown as a $1/3$ slice of a full pie chart. The scattering interactions from \textbf{a} have been marked (arrows and circles). \textbf{c}: Valence band ($\pi$) dispersion extracted along the $\Bar{\Gamma}-\Bar{\text{K}}-\Bar{\text{M}}$ high-symmetry directions, together with the extracted half-linewidth (in eV). The ARPES data has been overlaid with a simple nearest-neighbor tight binding calculation (dashed blue). Also marked are the measured energies for the VBM and the $\pi$-band van Hove (vH) singularity of hBN. The semi-transparent (blue) rectangle marks an interaction region of $\Delta E= 2\hbar\omega_{\text{D}}^{\text{BN}}$ around the vH energy.}
    \label{fig:fig2}
\end{figure}

The electronic structure of 8-layer hBN on graphene is mapped out at room temperature in Fig.~\ref{fig:fig2} for binding energies near the top of the $\pi$-band ($E_{\text{VBM}}$). The occupied bandstructure as shown was measured as a function of constant energy and simultaneously across all momentum vectors of the first Brillouin zone (BZ), using an aberration-corrected momentum microscope with its energy and momentum resolutions set to 50~meV and $0.02~\text{Å}^{-1}$, respectively \cite{Tusche2015spin,tusche2019imaging,Suppl_Mat}. The measured $\pi$-states reveal several signatures of electron scattering. Notably, a broadening appears along the $\Bar{\Gamma}-\Bar{\text{K}}$ direction at approximately $E_{\text{VBM}}+1$~eV (blue arrows) where also two faint and linearly dispersing features appear (grey arrow/circles), intersecting with the hBN $\pi$-band. Additional strong signs of broadening (green arrows) can be seen along the $\Bar{\text{K}}-\Bar{\text{M}}$ direction for energies closer to the valence band maximum (VBM). 

Sudden broadenings of the electronic structure like the ones mentioned are typical hallmarks of electron-boson interactions \cite{gayone2005determining,Hofmann2009electron,Valla1999many,LaShell2000non}. They signify a direct change in the quasiparticle lifetime $\tau$ -- or said differently, a reduction of the time an electron uses to fill a photohole \cite{Hofmann2009electron}. This phenomenon has already been thoroughly studied in hBN's sister compound graphene \cite{bostwick2007quasiparticle,Federico1}. Therein, spectroscopic signatures of lifetime renormalizations at large binding energies away from the Fermi level ($E_{\text{F}}$) have been reported \cite{Federico1}. Their origin has been debated\cite{jung2016sub}, but the evidence is pointing towards a strong electron-phonon coupling (EPC) in the $\sigma$-bands, mediated by the sudden onset of electron density of states (eDOS) at their band maxima \cite{Federico2,Hellsing2018phonon}. The EPC renormalizes the $\sigma$-bands with a large mass-enhancement (i.e. with a mass-enhancement parameter $\lambda\approx1$), manifesting itself as energy broadenings and `kinks' \cite{Federico2}. Additional broadenings near the $\sigma$- and $\pi$-band extremum points from phonon-mediated intra-band ($\pi\rightarrow\pi$) and inter-band ($\sigma\rightarrow\pi$) scattering have also been reported \cite{Hellsing2018phonon}. 

To ascertain the origins of the renormalizations observed from the hBN $\pi$-band, their magnitudes and binding energies were studied in more detail. In Fig.~\ref{fig:fig2}\textbf{c}, the measured $\Bar{\Gamma}-\Bar{\text{K}}-\Bar{\text{M}}$ wedge of the BZ is plotted along with the energy band half-linewidths ($\propto\tau^{-1}$) along these high-symmetry directions. The maximum broadening along $\Bar{\Gamma}-\Bar{\text{K}}$ appears at $\approx175$~meV lower binding energy than the $\pi$-band vH singularity at $\Bar{\text{M}}$. Interestingly, this energy separation matches roughly with the expected scattering energy $\hbar\omega_{\text{D}}^{\text{BN}}$ of the longitudinal optical phonon modes of hBN \cite{thingstad2020phonon,jung2015vibrational}. This is illustrated in Fig.~\ref{fig:fig2}\textbf{c} by the semi-transparent, blue interaction region overlaid on the measured $\pi$-bandstructure in this energy range. Given its demonstrated occurrence in graphene \cite{Hellsing2018phonon}, we postulate that the described renormalization comes from intra-band $\pi_{\text{BN}}$-$\pi_{\text{BN}}$ electron-phonon scatterings originating at the vH singularity ($\Bar{\text{M}}$) of hBN.

A more rigorous analysis of the observed energy renormalization is partially prevented by the faint, linear artifacts that intersect with the hBN $\pi$-band at a similar binding energy (blue arrows). Still, their effect is very local in the energy and momentum phase diagram. Naively, these artifacts can be mistaken for the $\pi$-band of the underlying graphene layer. However, the graphene $\pi$-band is only nearly linear close to the $\Bar{\text{K}}$ point, which is found at a radius of approximately $1.7~\text{Å}^{-1}$ relative to $\Bar{\Gamma}$ in both graphene and hBN \cite{roth2013chemical,Chen2017electronic}. Selective area photoemission measurements of the hBN flake and the adjacent graphene region showed that the two materials were more or less aligned rotationally on top of each other (within $\approx5^{\circ}$). Given the small misalignment, the linear features were situated at the wrong distance from $\Bar{\Gamma}$, and thus could not be a part of the graphene $\pi$-band itself \cite{Koch2018electronic,joucken2019nanospot}. 

Additionally, the true graphene $\pi$-band should cross the Fermi level \cite{bostwick2007quasiparticle}. However, a curvature analysis \cite{zhang2011precise} of the $\Bar{\Gamma}-\Bar{\text{K}}-\Bar{\text{M}}$ wedge from Fig.~\ref{fig:fig2}\textbf{c} revealed that the anomalous features were visible exclusively at binding energies larger than $E_{\text{VBM}}$, i.e., where the hBN has a finite eDOS (see the Supporting Information \cite{Suppl_Mat}). This signifies that the observed anomalies involve hBN-dependent transitions that are inelastic in energy and/or momentum. Based on similar observations in few-layer graphene, we infer that they are signatures of secondary electrons ejected from the limited number of unoccupied final states available \cite{strocov2000three,mahatha2011unoccupied,krivenkov2017suppression}. Alternatively, these could be from Umklapp-scattered electrons ejected from the underlying graphene, interacting with the hBN as they pass through the flaked material on their way out into vacuum \cite{ohta2012evidence}.

Similar to the case of the graphene $\sigma$-band, the strong eDOS increase set naturally by the hBN $\pi$-band maximum at the $\Bar{\text{K}}$ point is expected to enable phonon-mediated energy renormalizations \cite{Federico1,Federico2,thingstad2020phonon}. This is also corroborated by our complex self-energy $\Sigma$ analysis: along the $\Bar{\text{K}}-\Bar{\text{M}}$ direction an abrupt increase in half-linewidth near the VBM can be observed, being a typical hallmark of EPC \cite{gayone2005determining,Valla1999many,Hofmann2009electron}. At these energies the gradient of the $\pi$-band dispersion is small, and its measured energy half-linewidth translates to the imaginary part of the self-energy ($\Im\Sigma$) directly \cite{LaShell2000non,Hofmann2009electron,Mahatha2019electron}. The real part of the self-energy ($\Re\Sigma$) can be found from the discrepancy between the measured band position and the theoretically expected, non-interacting bandstructure:\\ $\Re\Sigma\equiv E(\mathbf{k})-\varepsilon(\mathbf{k})$ \cite{gayone2005determining}.

The non-interacting electronic band $\varepsilon(\mathbf{k})$ can be obtained using various approximations. For example, the $\pi$-band of mono- or multilayer hBN can be described by a tight-binding (TB) approximation, which is subsequently fitted to experimental results or first-principles calculations to obtain reasonable hopping parameters \cite{roth2013chemical,Koch2018electronic,ribeiro2011stability}. However, the resultant $\varepsilon(\mathbf{k})$ can only be used as an approximation at best, as the fitting will ignore all finite $\Re\Sigma$ contributions from any many-body interactions. Alternatively, the full bandstructure of the hBN-on-graphene system could be calculated from first principles, but this requires detailed knowledge of the hBN-substrate interaction and the stacking sequence and rotation of the hBN layers, which goes well beyond the scope of this work \cite{slawinska2010energy,Gilbert2019alternative}. 

\begin{figure}[t!]
    \centering
    \includegraphics[]{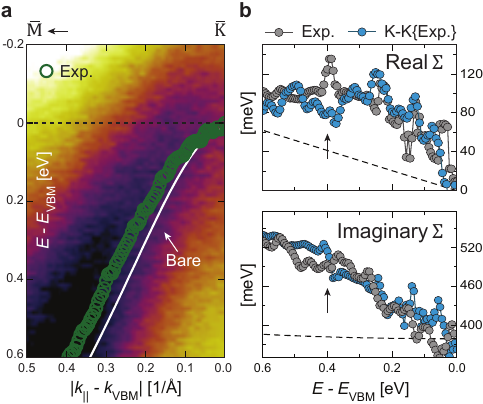}
    \caption{Energy renormalization near the hBN $\pi$-band maximum. \textbf{a}: The measured electronic structure of the hBN $\pi$-band, overlaid with the renormalized band position (green dots) as determined by curve fitting, and the unrenormalized band (white line) from a Kramers-Kronig analysis. \textbf{b}: The self-consistent $\Re\Sigma$ and $\Im\Sigma$ of the $\pi$-band corresponding to the spectrum shown in \textbf{a}. Both experimental $\Sigma$ contributions (gray) have been overlaid by a Kramers-Kronig transformed equivalent (blue), estimated from the opposite experimental component. An abrupt energy renormalization (arrows) on a background of electron-impurity and electron-electron interactions (dashed lines) can be distinguished from both $\Sigma$ components.}
    \label{fig:fig3}
\end{figure}

In general, discrepancies between the measured and calculated electronic structure will occur, and their magnitude will depend on the type and level of approximation used. A solution that allows us to circumvent this problem and properly establish the non-interacting `bare' energy band $\varepsilon(\mathbf{k})$, is to make no rigorous assumptions about its energy dispersion. Instead, the fact that $\Re\Sigma$ and $\Im\Sigma$ are causally related can be exploited, so that one is determined from the other via a Kramers-Kronig (K-K) transformation. This methodology has been previously adopted for graphene \cite{Federico1,Federico2}, and is discussed in detail in Refs.~\citenum{Kordyuk2005bare} and \citenum{Pletikosic2012finding}. A first guess for $\varepsilon(\mathbf{k})$ was found from a nearest-neighbor TB calculation using the parameters $t_{1}=2.92$~eV and $\Delta_{\text{BN}}=4.3$~eV \cite{robertson1984electronic,Shyu2014electronic}. Then, keeping the energy and momentum of the VBM fixed, the shape of $\varepsilon(\mathbf{k})$ was adjusted to satisfy causality between the $\Sigma$ components via the K-K transformation (see the Supporting Information for details \cite{Suppl_Mat}). The resulting, non-interacting $\pi$-band and $\Sigma$ components are shown in Fig.~\ref{fig:fig3}.

At a glance, a prominent energy renormalization can be readily distinguished from both the $\Re\Sigma$ and $\Im\Sigma$ near the VBM (arrows in Fig.~\ref{fig:fig3}\textbf{b}). Its characteristic functional shape and weakly increasing background towards larger binding energies indicate the presence of both electron-phonon and electron-electron interactions \cite{gayone2005determining,Valla1999many}. Indeed, had this feature originated from a crossing between hBN and the anomalous band features near the vH energy, it would have existed exclusively as a local broadening across the intersectional region \cite{ohta2012evidence,Hellsing2018phonon}. Instead, the observed energy renormalization persists at larger binding energies as expected \cite{Hofmann2009electron}.

\begin{figure}[t!]
    \centering
    \includegraphics[]{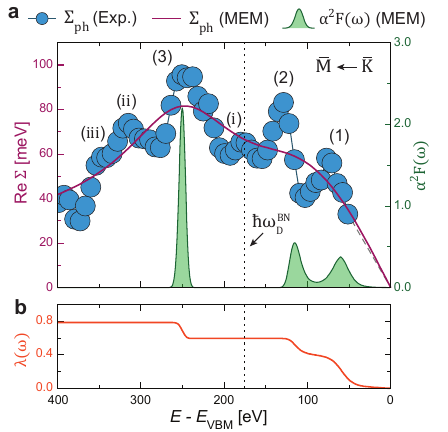}
    \caption{Signatures of electron-phonon interactions near the hBN $\pi$-band maximum. \textbf{a}: The experimentally determined, real self-energy contribution from electron-phonon interactions $\Re\Sigma_{\text{ph}}$ (in blue), overlaid with a best-fit model (purple line) and its corresponding Eliashberg function $\alpha^{2}F(\omega)$ (in green) as estimated by a MEM analysis. An experimental error $\sigma=5$~meV for each data point was estimated from the measured noise level and is shown by the size of the data point spheres. Additionally, the hBN Debye energy $\hbar\omega_{\text{D}}^{\text{BN}}\approx175$~meV has been indicated relative to the VBM (dashed vertical line). \textbf{b}: The integrated mass-enhancement factor $\lambda(\omega)$ of the electron-phonon coupling, estimated from the $\alpha^{2}F(\omega)$ shown in \textbf{a}.}
    \label{fig:fig4}
\end{figure}

A closer inspection of the measured $\Sigma$ components revealed a fine structure of multiple, distinct spectral signatures at well-defined binding energies. Hence, the $\Sigma_{\text{ph}}$ due to EPC was estimated from the data by subtracting the electron-impurity and electron-electron scattering contributions as shown in Fig.~\ref{fig:fig3}\textbf{b} (details in the Supporting Information \cite{Suppl_Mat}). The resulting $\Re\Sigma_{\text{ph}}$, with its leading features labeled, is shown in Fig.~\ref{fig:fig4}\textbf{a} (blue dots). We note that the renormalized and un-renormalized bands could not be unambiguously distinguished from one another at energies smaller than the instrument resolution (50~meV). Hence the first few data points of $\Re\Sigma_{\text{ph}}$ have been omitted \cite{Chien2015electron}.

From the resultant $\Re\Sigma_{\text{ph}}$, several distinct and peaked features can be readily distinguished. When referenced to the VBM, the first two appear at energies similar to the in-plane acoustic phonons of few-layer hBN \cite{jung2015vibrational}. A minor feature can be resolved at approximately the hBN Debye energy $\hbar\omega_{\text{D}}^{\text{BN}}$ (dashed line) \cite{thingstad2020phonon}. Surprisingly, three more features can also be seen at energies larger than $\hbar\omega_{\text{D}}^{\text{BN}}$, the first one at approximately $E_{\text{VBM}}+250$~meV.

To better understand the observed peaks and quantify their individual contributions to the overall EPC, their Eliashberg function $\alpha^{2}F(\omega)$ and corresponding electron mass-enhancement factors $\lambda_{n}$ were estimated. The former approximates the DOS of the interacting phonon modes present, and the latter quantifies the strengths of their coupling to the electrons \cite{Hofmann2009electron}. Using a maximum entropy method (MEM) procedure \cite{Shi2004direct,tang2004spectroscopic,Chien2015electron}, $\alpha^{2}F(\omega)$ was extracted from the data\cite{Suppl_Mat}, and the cumulative mass-enhancement from the different phonon modes estimated as $\lambda(\omega)=\sum_{n}\lambda_{n}=2\int_{0}^{\omega}\frac{\dd\omega'}{\omega'}\alpha^{2}F(\omega')$. The results, along with the corresponding best-fit to $\Re\Sigma_{\text{ph}}$, have been shown in Fig.~\ref{fig:fig4}. The energies and $\lambda_{n}$ of the electron-phonon interactions have been summarized in Table~\ref{tab:Couplings}.

Immediately, the two lowest energy features in $\alpha^{2}F(\omega)$ can be correlated to the peaks 1 and 2 as observed from $\Re\Sigma_{\text{ph}}$. Both signify a strong EPC in the acoustic energy regime, having mass-enhancement factors $\lambda_{1}=0.39$ and $\lambda_{2}=0.21$, respectively. Such strong coupling to the acoustic modes has already been predicted theoretically in hBN systems thicker than 1 ML \cite{slotman2014phonons}. Also, the observed, dominant feature (3) above the Debye energy is faithfully re-asserted with a $\lambda_{3}=0.19$. The smaller features ($i$-$iii$), although visible from the measured $\Re\Sigma_{\text{ph}}$, were not unambiguously resolved by the MEM analysis when accounting for the thermal broadening of the measurements.

\begin{table}[t]
    \centering
    \caption{A summary of the peak assignments from Fig.~\ref{fig:fig4}, along with their energies, contributions to $\lambda$, and suggested scattering phonons. The entries in parenthesis are distinguishable from $\Re\Sigma$ but were not resolved by the MEM analysis. The phonon mode suggestions have been based on their measured and calculated energy values from Refs.~\citenum{jung2015vibrational} and \citenum{senga2019position}.}
    \label{tab:Couplings}
    \vspace{0.1cm}
    \begin{tabular}{clll}
        \hline
        Peak & $\hbar\omega$ [meV] & $\lambda_{n}$ & Phonon mode(s) \\
        \hline
        1 & $60\pm12$ & 0.39 & TA$_{\text{BN}}$/ZO$_{\text{BN}}$ \\
        2 & $115\pm9$ & 0.21 & TA$_{\text{BN}}$/ LA$_{\text{BN}}$/ZO$_{\text{BN}}$ \\
        ($i$) & $175\pm16$ & minor & TO$_{\text{BN}}$/ LO$_{\text{BN}}$ \\
        3 & $245\pm5$ & 0.19 & $2\times\text{LA}_{\text{BN}}$ \\
        ($ii$) & $315\pm16$ & minor & $2\times \text{TO}_{\text{BN}}$\ \\
        ($iii$) & $345\pm13$ & minor & $2\times \text{LO}_{\text{BN}}$ \\
        \hline
    \end{tabular}
\end{table}

Based on their energies, the observed peaks above the Debye energy cannot be explained by single-phonon scattering alone. Their presence, however, hints at the possibility of multi-phonon scattering events. Consecutive electron scattering by multiple phonons has been predicted from theory, suggesting that substantial scattering rates may occur in insulating and semiconducting polar materials \cite{Vogl1976microscopic,lee2020ab}. Additionally, experimental signatures of multi-phonon scattering in few-layer hBN have been suggested from inelastic electron tunneling spectroscopy (IETS) measurements \cite{jung2015vibrational}. However, a mode-specific quantification of any multiple-phonon interactions with electrons has, until now, not been presented. From the known phonon energies of hBN \cite{jung2015vibrational,senga2019position} we are able to suggest different two-phonon combinations to explain the observed higher-order peaks above the Debye energy. These, along with suitable phonon modes for the EPCs at lower energies, have also been summarized in Table~\ref{tab:Couplings}.

While electron-phonon scattering between the hBN and the underlying graphene cannot be ruled out completely, we can render it unlikely in the present case. Previous ARPES studies of graphene-on-hBN have indicated that coupling of the two materials should lead to large (i.e., Fr\"{o}hlich) polaron formation at the interface\cite{chen2018emergence}. In contrast, no such signatures of large polaron formation can be observed here \cite{Suppl_Mat,franchini2021polarons}. Examining other exfoliated hBN flakes with different thicknesses and rotational alignment with the substrate led to the same conclusion. Alternatively, one could suggest that the EPC peaks presented in Table~\ref{tab:Couplings} resulted from electron-phonon scattering between the hBN and graphene bands \cite{Hellsing2018phonon}. However, inter-material EPC for the current system is expected to be weak when compared to the theoretical EPC of few-layer hBN \cite{slotman2014phonons}, and the mass-enhancement demonstrated from our analysis.

We conclude that the apparent correspondence between the measured EPCs and scattering with the graphene is coincidental. We reiterate, however, that our MEM analysis was unable to properly distinguish all of the engaging hBN phonon modes. This was in part caused by the achievable instrumental resolution, but primarily by the thermal broadening at room temperature. For instance, further investigating the potential coupling to the lowest-energy  modes using a more sensitive method, e.g.~helium atom scattering\cite{holst2021material,benedek2021electron}, could provide additional insight. However, this would require larger-area single crystalline samples to be achieved.

In summary, we have reported the existence of several EPC processes that renormalize the $\pi$-band electronic structure of hBN. Together, they cause an increase in the electron scattering rate and a decrease in energy state lifetime. This may have severe implications for the electron transport in any hBN-adjacent conducting layers, e.g. in vdW heterostructures. Thus, our work not only sheds light on the complex many-body ground state of few-layer hBN; it also provides valuable insight into the possible scattering mechanisms that may hamper the performances of hBN-based electronic devices.

\section{Supporting Information}
Further details on the hBN sample preparation, the measurement parameters, the curvature analysis of the hBN $\pi$-states, and the interpretation of the measured self-energies.

\section{Acknowledgements}
The authors acknowledge financial support from the Research Council of Norway (RCN) through project no. 324183, 315330, 262633, 280788, and 245963. K.W. and T.T. also acknowledge support from JSPS KAKENHI (Grant Numbers 19H05790, 20H00354, and 21H05233). Part of the research was performed at the ESM 21-ID-2 beamline of the National Synchrotron Light Source II, a U.S. Department of Energy (DOE) Office of Science User Facility operated for the DOE Office of Science by Brookhaven National Laboratory under Contract number DE-SC0012704. This work also used the resources of the Center for Functional Nanomaterials, Brookhaven National Laboratory, which is supported by the U.S. Department of Energy, Office of Basic Energy Sciences, under Contract number DE-SC0012704. We would also like to thank T. Balasubramanian, T.-Y. Chien, A. Grubi\v{s}i\'{c}-\v{C}abo, A. Ettema, T. Frederiksen, B. Hellsing, J. Manson, D. Pohlenz, A. Qaiumzadeh, A. Sudbø, and E. Thingstad for fruitful discussions.

\providecommand{\latin}[1]{#1}
\makeatletter
\providecommand{\doi}
  {\begingroup\let\do\@makeother\dospecials
  \catcode`\{=1 \catcode`\}=2 \doi@aux}
\providecommand{\doi@aux}[1]{\endgroup\texttt{#1}}
\makeatother
\providecommand*\mcitethebibliography{\thebibliography}
\csname @ifundefined\endcsname{endmcitethebibliography}
  {\let\endmcitethebibliography\endthebibliography}{}

\begin{figure*}
    \centering
    \includegraphics[width=8.25cm]{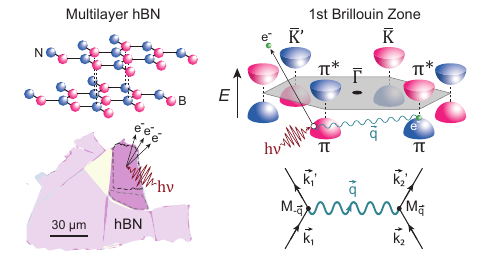}
    \caption{TOC graphic}
\end{figure*}

\end{document}


\title{Supporting Information:\\Phonon-Mediated Quasiparticle Lifetime Renormalizations in Few-Layer Hexagonal Boron Nitride}


\author{H\r{a}kon I. R\o{st}} \email{hakon.rost@uib.no}
\affiliation{\footnotesize{Department of Physics and Technology, University of Bergen, All\'egaten 55, 5007 Bergen, Norway.}}
\affiliation{\footnotesize{Department of Physics, Norwegian University of Science and Technology (NTNU), NO-7491 Trondheim, Norway.}}

\author{Simon P. Cooil}
\affiliation{\footnotesize{Department of Physics and Centre for Materials Science and Nanotechnology, University of Oslo (UiO), Oslo 0318, Norway.}}

\author{Anna Cecilie Åsland}
\author{Jinbang Hu}
\affiliation{\footnotesize{Department of Physics, Norwegian University of Science and Technology (NTNU), NO-7491 Trondheim, Norway.}}

\author{Ayaz Ali}
\affiliation{\footnotesize{Department of Smart Sensor Systems, SINTEF DIGITAL, Oslo, 0373, Norway.}}
\affiliation{\footnotesize{Department of Electronic Engineering, Faculty of Engineering \& Technology, University of Sindh, Jamshoro, 76080, Pakistan.}}

\author{Takashi Taniguchi}
\affiliation{\footnotesize{International Center for Materials Nanoarchitectonics, National Institute for Materials Science, 1-1 Namiki, Tsukuba 305-0044, Japan.}}

\author{Kenji Watanabe}
\affiliation{\footnotesize{Research Center for Functional Materials, National Institute for Materials Science, 1-1 Namiki, Tsukuba 305-0044, Japan.}}

\author{Branson D. Belle}
\affiliation{\footnotesize{Department of Smart Sensor Systems, SINTEF DIGITAL, Oslo, 0373, Norway.}}

\author{Bodil Holst}
\affiliation{\footnotesize{Department of Physics and Technology, University of Bergen, All\'egaten 55, 5007 Bergen, Norway.}}

\author{Jerzy T. Sadowski}
\affiliation{\footnotesize{Center for Functional Nanomaterials, Brookhaven National Laboratory, Upton, New York, 11973, USA.}}

\author{Federico Mazzola}
\affiliation{\footnotesize{Department of Molecular Sciences and Nanosystems, Ca’ Foscari University of Venice, 30172 Venice, Italy.}}
\affiliation{\footnotesize{Istituto Officina dei Materiali, Consiglio Nazionale delle Ricerche, Trieste I-34149, Italy.}}

\author{Justin W. Wells} \email{j.w.wells@fys.uio.no}
\affiliation{\footnotesize{Department of Physics, Norwegian University of Science and Technology (NTNU), NO-7491 Trondheim, Norway.}}
\affiliation{\footnotesize{Department of Physics and Centre for Materials Science and Nanotechnology, University of Oslo (UiO), Oslo 0318, Norway.}}

\maketitle
\vspace{-1cm}
\section{Sample Preparation}\label{subsec:sampleprep}
\begin{figure}[t]
    \centering
    \includegraphics[]{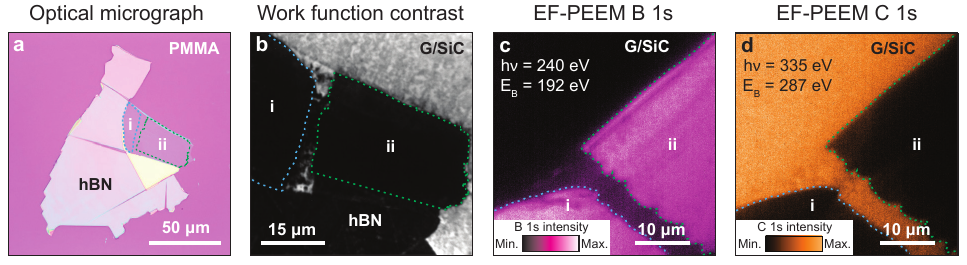}
    \caption{Preparation and characterization of a few-layer hBN sample. \textbf{a}: Optical micrograph of the exfoliated hBN flake on PMMA that was selected for further transfer onto graphene on SiC. \textbf{b}: The same exfoliated flake measured after the transfer onto graphene on SiC, using work function contrast PEEM near the secondary electron cutoff ($h\nu\approx5.2$~eV). The regions $i$ and $ii$ encompassed by the dashed, blue and green lines in both subfigures were the ones used for concomitant LEEM and PEEM studies. \textbf{c}, \textbf{d}: Energy-filtered PEEM micrographs of the exfoliated hBN flake, performed at the binding energies $E_{\text{B}}$ of the B~1s and C~1s core levels. The B~1s signal predominantly occurs within regions $i$ and $ii$, whereas the C~1s signal can only be observed from the graphene substrate.}
    \label{fig:figS1}
\end{figure}

Crystals of bulk hexagonal boron nitride (hBN) were prepared according to a `temperature-gradient method' at high pressure and temperature, using barium boron nitride (\ce{Ba3B2N4}) as a solvent system \cite{watanabe2004direct}. Thin flakes of hBN were micromechanically exfoliated \cite{blake2007making} from the bulk crystals and transferred onto a PMMA/PVA/SiO$_2$ stack. Suitable samples of few-layer hBN on the polymer stack were identified by optical contrast microscopy (Fig.~\ref{fig:figS1}\textbf{a}), and subsequently transferred onto substrates of monolayer graphene on 6H-SiC(0001) \cite{yu2011new}, using a dry transfer method at \SI{80}{\celsius} \cite{dean2010boron}. 
Excess polymer from the transfer was removed by annealing the hBN/graphene/SiC heterostructures under ultrahigh vacuum (pressure~$\leq1\times10^{-9}$~mbar) to \SI{300}{\celsius} for a few hours, followed by several flashes to \SI{600}{\celsius} for short durations.

\section{LEEM and PEEM Measurements}
LEEM, $\mu$-LEED, and XPEEM measurements were performed at the ESM 21-ID-2 (LEEM/PEEM) beamline of National Synchrotron Light Source II (NSLS-II), using an Elmitec aberration-corrected LEEM/PEEM. The presence and position of a suitable
and properly grounded exfoliated hBN flake on the graphene substrate were ascertained by LEEM and spatially resolved, surface-sensitive XPEEM measurements of the B~1s and C~1s core levels (Figs.~\ref{fig:figS1}\textbf{c} and \ref{fig:figS1}\textbf{d}, inelastic mean-free path $<\SI{0.5}{\nm}$). Additionally, dispersive plane XPS measurements of the B~1s, C~1s, and O~1s were measured within the region of the flake to ascertain the composition of the sample. The low-energy electron reflectivity (LEER) imaging measurements (discussed in the main text) were performed in the range 0-10~eV, using an incident and coherent electron beam. The $\mu$-LEED was collected using a \SI{1.5}{\um} diameter selective area aperture placed within the region of the hBN flake.

Further photoemission measurements in both real and $k$-space were performed using an aberration-corrected, energy-filtered photoemission electron microscope (EF-PEEM) working at an extraction voltage \SI{12}{\kilo\volt} (NanoESCA III, Scienta Omicron GmbH). Real-space imaging was performed with a pass energy $E_{\text{P}}=\SI{50}{\eV}$, using a \SI{150}{\um} contrast aperture, a \SI{0.5}{\milli\metre} entrance slit, and a non-monocromated Hg source ($h\nu\approx\SI{5.2}{\eV}$) for photoexcitation. With these settings, the microscope had the nominal real space and energy resolutions $\Delta x=\SI{35}{\nm}$ and $\Delta E =\SI{100}{\milli\eV}$, respectively. The hBN bandstructure was reconstructed from a series of measured constant energy surfaces (CES, $E$ vs. $k_{\text{x}}$, $k_{\text{y}}$) obtained in the diffractive plane mode of the same momentum microscope \cite{Tusche2015spin,tusche2019imaging}. The bandstructure measurements were restricted to an area within the hBN flake by introducing an iris aperture of $\approx30\times\SI{15}{\um}^{2}$. Using a He I photoexcitation source ($h\nu=\SI{21.22}{\eV}$), CES were acquired at $E_{\text{P}}=\SI{25}{\eV}$ with a \SI{0.5}{\mm} entrance slit to the energy filter, yielding energy and momentum resolutions of $\Delta E =\SI{50}{\meV}$ and $\Delta k = 0.02~\text{Å}^{-1}$, respectively. Each CES was acquired $\SI{10}{\milli\eV}$ apart and with a $k$-space field of view of $4.3~\text{Å}^{-1}$, i.e. spanning beyond the 1st Brillouin zone of hBN. All measurements were performed at room temperature ($\approx\SI{300}{\kelvin}$). 

\begin{figure}[t]
    \centering
    \includegraphics[]{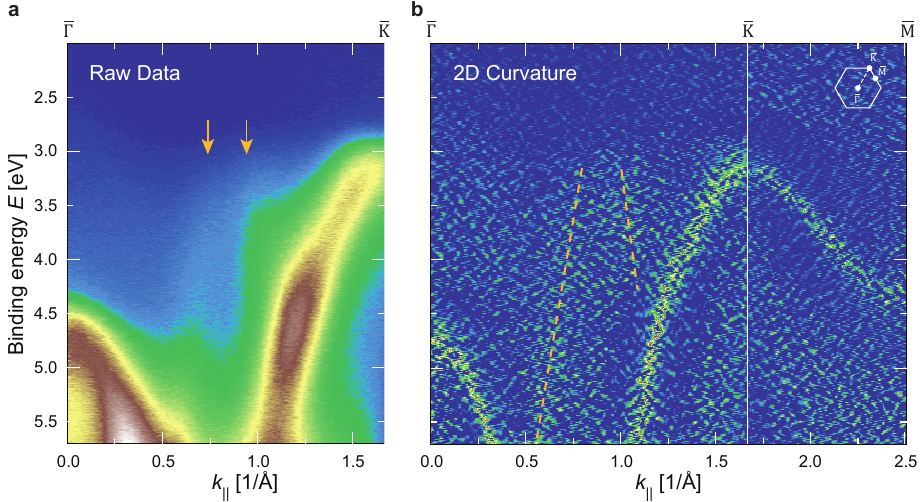}
    \caption{Curvature analysis of the $\bar{\Gamma}-\bar{\text{K}}-\bar{\text{M}}$ wedge. The positions of the anomalous, near-linearly dispersing bands have been highlighted by arrows (\textbf{a}) and dashed lines (\textbf{b}). From the two-dimensional curvature plot (\textbf{b}), the anomalous bands diminish in intensity at binding energies above the hBN valence band maximum.}
    \label{fig:figS2}
\end{figure}

\vspace{-0.5cm}
\section{Curvature Analysis of $\bar{\Gamma}-\bar{\text{K}}-\bar{\text{M}}$}
To assess the presence, shape, and intensity of the anomalous features observed from the hBN ARPES measurements, a two-dimensional (2D) curvature analysis of the $\bar{\Gamma}-\bar{\text{K}}-\bar{\text{M}}$ wedge was performed (see Ref.~\onlinecite{zhang2011precise} for details). The raw data and its resulting 2D curvature plot are shown in Fig.~\ref{fig:figS2}. The near-linearly dispersing anomalous features can be distinguished near halfway along $\bar{\Gamma}-\bar{\text{K}}$, as indicated by arrows in Fig.~\ref{fig:figS2}\textbf{a}. Notably, their intensity diminishes at binding energies above the hBN valence band maximum as shown in Fig.~\ref{fig:figS2}\textbf{b}. Also, no energy-shifted replicas of the hBN $\pi$-band from polaron formation can be distinguished from the 2D curvature analysis \cite{chen2018emergence,franchini2021polarons}.

\vspace{-0.5cm}
\section{Rotational Averaging of the Electronic Structure}
Before extracting anything from the measured bandstructure, the three-dimensional data cone ($k_{x}$, $k_{y}$, $E$) was replicated $n=6$ times. Each copy was rotated around the energy axis, i.e. the $\Bar{\Gamma}$ point of the BZ, by $\varphi=n\pqty{2\pi/6}$ before adding them all together. This was done to minimize any geometrically induced differences in the photoionization matrix elements \cite{day2019computational}, and furthermore to improve the signal-to-noise ratio of the measured bands.

\vspace{-0.5cm}
\section{Approximating the Non-Interacting $\pi$-Band}
Energy distribution curves (EDCs) of the $\pi$-band were fitted over the relevant energy range using a Lorentzian line shape, approximating the background by an error function to accommodate the visible `step' in intensity near the VBM. The band's peak position and linewidth (in eV) were extracted and used to estimate the $\Re\Sigma$ and $\Im\Sigma$, respectively, using an analysis procedure similar to the ones demonstrated in Refs.~\onlinecite{Kordyuk2005bare,Pletikosic2012finding,Mazzola2017strong,mazzola2022tuneable}. As an initial guess for the unperturbed $\pi$-band, a nearest-neighbor tight-binding (TB) calculation was performed with $t_{1}=\SI{2.92}{\eV}$ and $\Delta_{\text{BN}}=\SI{4.3}{\eV}$ \cite{robertson1984electronic,Shyu2014electronic}. The initial bare band curve obtained from the TB calculation was approximated by a fifth-degree polynomial over the same energy range as the fitted experimental band. Finally, its shape was iteratively adjusted to achieve consistency between the measured, real and imaginary parts of the self-energy ($\Sigma_{\text{exp}}$) through the Kramers-Kronig transform \cite{Kordyuk2005bare,Pletikosic2012finding}.

\vspace{-0.5cm}
\section{Estimating the Inelastic Background of $\Re\Sigma$ and $\Im\Sigma$}
To estimate the measured inelastic energy losses due to non-bosonic interactions, a simplistic model $\Sigma$ was constructed and fitted to the measured $\Sigma_{\text{exp}}$. The model consisted of a linear combination of contributions describing the electron-impurity ($\Sigma_{\text{el-imp}}$), electron-electron ($\Sigma_{\text{el-el}}$), and electron-phonon ($\Sigma_{\text{el-ph}}$) scattering. The free parameters of the model were in turn adjusted to fit the measured $\Im\Sigma_{\text{exp}}$ by minimizing their root mean square (RMS) difference, and subsequently, Kramers-Kronig transformed to find the model $\Re\Sigma$. Each fitting parameter uncertainty was
estimated from a relative RMS increase of $\pm5\%$.

The term $\Im\Sigma_{\text{el-imp}}$ was modeled as a constant broadening of $380~\text{meV}$ ($\propto T$), estimated from the full width at half maximum (FWHM) of the fitted EDC at the $\Bar{\text{K}}$ point. We note here that using the FWHM directly will overestimate $\Im\Sigma_{\text{imp}}$, as it also contains additional, constant energy broadening from the instrument's finite energy resolution. The term $\Im\Sigma_{\text{el-el}}$ was modeled as a logarithmically corrected, quadratic expression for a two-dimensional Fermi liquid \cite{higashiguchi2007high,hufner2007very}:
\begin{equation}\label{eq:elel}
\beta\cdot [E-E_{\text{VBM}}]^{2}\ln|E-E_{\text{VBM}}|,
\end{equation}
with $\beta=3.0\pm0.3~\text{eV}^{-1}$.

Finally, $\Im\Sigma_{\text{el-ph}}$ was modeled by a sum of contributions $\Im\Sigma_{\text{el-ph}}^{(i)}$ ($i=1,2$) from two separate phonon modes with different maximum energies. The density of states $F^{(i)}(\omega)$ of each phonon mode was described using a two-dimensional and isotropic Debye model ($\propto\omega$), having a distinct and characteristic maximum (cutoff) frequency $\omega_{i}$. Using Eliashberg formalism, each contribution ($i$) to $\Im\Sigma_{\text{el-ph}}$ from the individual electron-phonon interactions was calculated as \cite{hufner2007very,Hellsing2002electron}:
\begin{equation}\label{eq:Eliashberg}
    \Im\Sigma_{\text{el-ph}}^{(i)}\pqty{\omega,T}= \pi\int_{0}^{\omega_i}\alpha^{2}F^{(i)}\pqty{\omega'}\cdot[1+2n\pqty{\omega',T}
    +f\pqty{\omega+\omega',T}-f\pqty{\omega-\omega',T}]\dd\omega',
\end{equation}
where $n\pqty{\omega,T}$ and $f\pqty{\omega,T}$ are the boson and fermion distributions, respectively. The term $\alpha^{2}F^{(i)}\pqty{\omega}=(\lambda_{\text{ph}}^{(i)}/2)\pqty{\omega/\omega_i}$ for $\omega<\omega_i$, where $\lambda_{\text{ph}}^{(i)}$ is the dimensionless mass enhancement factor for coupling to the phonon mode $i$. Above each cut-off frequency $\omega_i$, the corresponding $\alpha^{2}F^{(i)}\pqty{\omega}=0$. The optimized two-phonon Eliashberg model yielded peaks at energies $\hbar\omega_1=185\pm35$~meV and $\hbar\omega_2=335\pm15$~meV below the VBM, with the mass-enhancement factors $\lambda_1=0.18\pm0.04$ and $\lambda_2=0.36\pm0.03$, respectively.

\vspace{-0.5cm}
\section{Estimating the Electron-Phonon Coupling from $\Re\Sigma$}
The real counterpart to the $\Im\Sigma_{\text{el-ph}}$ expression in Eq.~\ref{eq:Eliashberg} is defined as \cite{Hofmann2009electron}:
\begin{equation}\label{eq:ReSE}
    \Re\Sigma_{\text{el-ph}}\pqty{\omega,T} = \int_{-\infty}^{\infty}\dd\nu\int_{0}^{\omega}\alpha^{2}F\pqty{\omega'}\cdot\frac{2\omega'}{\nu^{2}-\omega'^{2}}\cdot f\pqty{\nu+\omega,T}\dd\omega'.
\end{equation}
Here, $\alpha^{2}F\pqty{\omega}$ is the total phonon density of states (DOS) weighted by the effective strength $\alpha^{2}$ of the electron-phonon coupling. Thus, it describes the phonon modes participating in the interaction and contributing to $\Re\Sigma_{\text{el-ph}}$. From $\alpha^{2}F\pqty{\omega}$, the total mass-enhancement factor $\lambda$ of the electrons can be found by \cite{grimvall1981electron}:
\begin{equation}\label{eq:lambda}
    \lambda_{\text{ph}} = 2\int_{0}^{\omega_{\text{max}}}\frac{\alpha^{2}F\pqty{\omega'}}{\omega'}\dd\omega',
\end{equation}
where $\omega_{\text{max}}$ is the maximum observable phonon energy. If $\Re\Sigma_{\text{el-ph}}$ is known, e.g., from ARPES measurements, then $\alpha^{2}F\pqty{\omega}$ can be estimated from the expression in Eq.~\ref{eq:ReSE} by integral inversion. The result can then be used to predict which (combinations of) phonon modes participate, and their interaction strength $\lambda_{\text{ph}}$ with the electrons can be found from Eq.~\ref{eq:lambda}. 

To overcome the challenge of mathematical instability posed by the direct integral inversion of $\Re\Sigma_{\text{el-ph}}$ a maximum entropy method (MEM) analysis procedure was employed. A detailed outline of the method and its theoretical foundations can be found in Refs.~\onlinecite{Gubernatis1991quantum,Shi2004direct,tang2004spectroscopic}. Therefore, only a brief summary is given here. The optimum approximation to the experimental $\Re\Sigma_{\text{el-ph}}$ is obtained by minimizing the functional:
\begin{equation}\label{eq:MEMfunctional}
    L = \frac{\chi^2}{2} - aS,
\end{equation}
where $\chi^2$ is the standard deviation between the measured and calculated $\Re\Sigma_{\text{el-ph}}$. The term $aS$ serves to prevent overfitting of the experimental data by imposing physical constraints on how the $\alpha^{2}F\pqty{\omega}$ term should look. $S$ is the Shannon-Jaynes entropy \cite{Shi2004direct}:
\begin{equation}\label{eq:entropy}
    S = \int_{0}^{\infty}\dd\omega'\bqty{\alpha^{2}F\pqty{\omega'} - m\pqty{\omega'} - \alpha^{2}F\pqty{\omega'}\ln{\frac{\alpha^{2}F\pqty{\omega'}}{m\pqty{\omega'}}} },
\end{equation}
with the constraint model $m\pqty{\omega}$ reflecting our \emph{a priori} knowledge of $\alpha^{2}F\pqty{\omega}$. The term $a$ is a coefficient in the conditional probability of $\alpha^{2}F\pqty{\omega}$ under the constraints set by $m\pqty{\omega}$ \cite{skilling1989maximum}. In `classical' MEM, the probability defined by $a$, $S$ and $m\pqty{\omega}$ is maximized to minimize $L$ \cite{Gubernatis1991quantum}.

For the analysis of the hBN $\pi$-band, the constraint model was chosen to be:
\begin{equation}\label{eq:constraints}
    m\pqty{\omega}=
    \begin{cases}
    m_{0}\pqty{\omega/\omega_{\text{A}}}^{2}, & \text{if } 0\leq\omega<\omega_{\text{A}}, \\
    m_{0}, & \omega_{\text{A}}\leq\omega\leq\omega_{\text{max}}, \\
    0, & \text{otherwise}.
    \end{cases} \\
\end{equation}
This model reflects the expected quadratic increase in DOS up to the maximum energy $\hbar\omega_{\text{A}}$ of the first distinguishable acoustic mode \cite{kittel2021introduction}. The constant $m_{0}$ is positive, thus making $m\pqty{\omega}$ positive definite over the relevant energy range. The energy $\hbar\omega_{\text{max}}$ defines the cutoff energy for the observed electron-phonon interactions.

Initially, $m_{0}$ was set to approximately the average value expected for $\alpha^{2}F\pqty{\omega}$. The energy parameter $\hbar\omega_{\text{A}}$ was set to approximately the $E-E_{\text{VBM}}$ value where the first peak in the experimental $\Re\Sigma_{\text{ph}}$ could be observed (see Fig.~4\textbf{a} in the main text). Finally, $\hbar\omega_{\text{max}}$ was fixed at $E-E_{\text{VBM}}=400$~meV at the step-like cut-off in the experimental $\Re\Sigma_{\text{ph}}$ and $\Im\Sigma_{\text{ph}}$ (Fig.~3\textbf{b}, main text). In minimizing $L$, the parameters $m_{0}$ and $\hbar\omega_{\text{A}}$ were iteratively adjusted, achieving the final values $\hbar\omega_{\text{A}} = 60$~meV and $m_{0}=0.35$. This resulted in  a satisfying fit to $\Re\Sigma_{\text{ph}}$, with the constraint function recreating the main feats of $\alpha^{2}F\pqty{\omega}$ (i.e., positive definite, initial quadratic increase). Yet, it was still sufficiently structureless for an unbiased fitting of the data \cite{Shi2004direct}.

\newpage